\documentclass[letterpaper,twocolumn,10pt]{article}
\usepackage{usenix2019_v3}
\usepackage[utf8]{inputenc}
\usepackage[utf8]{inputenc}
\usepackage{appendix}
\usepackage{filecontents}
\usepackage{cite}
\usepackage{subcaption}
\usepackage{amsmath,amssymb,amsfonts}
\usepackage{algorithmic}
\usepackage{graphicx}
\usepackage{textcomp}
\usepackage[colorinlistoftodos]{todonotes}
\usepackage{bmpsize}
\usepackage{xspace}
\usepackage{lipsum}
\usepackage{pgfplots}
\pgfplotsset{compat=1.14}
\usepackage[official]{eurosym}

\usepackage{versions}
\excludeversion{RedundantContent}   
\excludeversion{Study2}
\usepackage{multirow}
\usepackage{booktabs}

\usepackage{subcaption}
\usepackage{float}

\begin{document}
\pagestyle{empty}
\date{}

\title{\Large \bf ``I feel invaded, annoyed, anxious and I may protect myself”: Individuals’ Feelings about Online Tracking and their Protective Behaviour across Gender and Country} 

\author{
{\rm Kovila P.L. Coopamootoo}\\
Newcastle University, UK\\
kovila.coopamootoo@newcastle.ac.uk
\and
{\rm Maryam Mehrnezhad}\\
Newcastle University, UK\\
maryam.mehrnezhad@newcastle.ac.uk
\and
{\rm Ehsan Toreini}\\
Durham University, UK\\
ehsan.toreini@durham.ac.uk
} 

\maketitle

\begin{abstract}
Online tracking is a primary concern for Internet users, yet previous research has not found a clear link between the cognitive understanding of tracking and protective actions.
We postulate that protective behaviour follows affective evaluation of tracking.
We conducted an online study, with N=614 participants, across the UK, Germany and France, to investigate how users feel about third-party tracking and what protective actions they take. 
We found that most participants' feelings about tracking were negative, described as deeply intrusive - beyond the informational sphere, including feelings of annoyance and anxiety, that predict protective actions.
We also observed indications of a `\emph{privacy gender gap}', where women feel more negatively about tracking, yet are less likely to take protective actions, compared to men.
And less UK individuals report negative feelings and protective actions, compared to those from Germany and France.
This paper contributes insights into the affective evaluation of privacy threats and how it predicts protective behaviour.
It also provides a discussion on the implications of these findings for various stakeholders, make recommendations and outline avenues for future work.
\end{abstract}

\begin{RedundantContent}

This paper reports Internet users' strong disapproval of TPT, which can act as a call for more privacy-supportive advertising business models. 
It also contributes research into how protective actions follow emotional evaluations, and expand literature on awareness and use of PETs.
Our findings support arguments that women are vulnerable online, and in need of support.
\end{RedundantContent}

\section{Introduction}
In recent years, internet advertising has become increasingly tailored to individual users, and is often referred to as online behavioural advertising or targeted advertising~\cite{englehardt2016online}. 
When users visit a web page, its contents can come from a first- or third-party, where the first-party is the one the user is explicitly visiting, while the third-party includes advertising networks, analytics companies and social networks that contract with first-party websites~\cite{mayer2012third}. The first- or third-party places cookies as tracking mechanism on users' devices.
In addition, online tracking has gradually developed into more sophisticated methods to exploit user data, via browser-based fingerprinting~\cite{pugliese2020long} and tracking across multiple devices, where fingerprinting information combined with cookies can provide a well-targeted data collection and tracking of users. 
Other tracking technologies include web beacons, clear GIFs, page tags and web bugs, that take the form of a small, transparent image. 
In European countries, the General Data Protection Regulation (GDPR)~\cite{GDPR16} and the proposed ePrivacy Regulation~\cite{ePrivacyRegulation},
with the UK establishing its implementation~\cite{parliament2018data}, stipulate that
online service providers are required to inform individuals that tracking technologies are present, what they do and why, and to receive users' consent to use them. 


The general public opinion in national surveys as reported in the UK and Europe is that tracking online is a privacy concern for most citizens~\cite{blank2019perceived,european2015special}, and objecting to receiving direct marketing is the most exercised right in Europe (accounting for $24\%$ of participants surveyed in 2019)~\cite{eurobarometer2019european}.
However, research shows that individuals have inaccurate and incomplete mental models of the mechanics of behavioural advertising~\cite{yao2017folk} and have misconceptions about the purpose of cookies~\cite{mcdonald2010beliefs}. 
They perceive behavioural advertising to be either (or simultaneously) useful in providing ads that match their interests~\cite{ur2012smart,mcdonald2010beliefs,auxier2019americans}, 
privacy-invasive~\cite{ur2012smart, mcdonald2010beliefs,turow2009americans}, 
creepy~\cite{ur2012smart}, embarrassing or suggestive~\cite{agarwal2013not},
and often show varying acceptance of behavioural advertising depending on the context~\cite{chanchary2015user,wang2016examining,melicher2016not}.
With regards to the use of privacy technologies for tracking protection, users' mental models of tracking (that is their cognitive evaluation of tracking) only weakly relate to their use of tracker-blocking extensions~\cite{mathur2018characterizing}, where tracking protection tools are also thought to suffer from usability issues~\cite{schaub2016watching,leon2015privacy,ValuePRivacy}.

We postulate instead that \emph{affective evaluations} can better predict protective actions.
We ground our proposition in literature on the informative and decisive value of emotions, 
where models such as the affect-as-information hypothesis~\cite{clore2001affect}, the risk-as-feelings hypothesis~\cite{loewenstein2001risk}, or the use of affective heuristics when judging risky situations~\cite{slovic2002rational}, all explain the influence of affect in decisions and on behaviour.


In addition to all Internet users potentially facing threats to their privacy, 
demographic and personal attributes may influence the experience of tracking and protective behaviour.
In particular, it is known that women tend to be more sensitive and concerned about their privacy online compared to men~\cite{oomen2008privacy,buchi2017caring}, 
who have been shown to be more familiar with various privacy protection methods and to use them more often than women~\cite{oomen2008privacy}.
Furthermore, individuals of different countries may 
use different privacy protection practices~\cite{coopamootoo2020usage}, where
variation in behaviour may be due to differences in the importance attributed to data protection~\cite{soffer2014privacy}, in sensitivity with regards to the duration or quantity of data collected~\cite{cvrcek2006study} or in the perceived risks of privacy violations~\cite{krasnova2010privacy}.

As a consequence, we aim to investigate how feelings about (third-party) tracking associate with and predict protective behaviour, across gender and country. To enable this inquiry, we define the following research questions, starting with elicitation of feelings and protective actions: 

\textbf{RQ1} ``\emph{How do individuals feel with regards to third-party tracking?}", given their gender and country differences.

\textbf{RQ2} ``\emph{What tracking protective actions do individuals employ online?}", given their gender and country differences.

\textbf{RQ3}: ``\emph{How are individuals' feelings about third-party tracking associated with their protective actions?}"

\textbf{RQ4}: ``\emph{How do individuals’ feelings about third-party tracking predict whether they take protective actions or not, given their gender and country differences?}"

\emph{Contributions.} 
This paper contributes a relatively large-scale, gender-based and cross-national investigation of user tracking protection, to the rich user-centric privacy behaviour and tracking protection research area. 
It employs qualitative and quantitative analyses to answer the research questions.
We summarise our findings as follows:

(1) most individuals ($71.8\%$ of our sample) feel negatively about third-party tracking, where feeling tones can be broadly categorised into (a) generally not okay/negative, (b) sometimes okay, sometimes not okay, (c) generally okay/indifferent, or (d) other tones;

(2) individuals employ tracking protection actions that can be grouped within $9$ technology types and a relatively large \% do not take any protective actions ($34.7\%$ of our sample);

(3) there is a significant association between feelings about third-party tracking and self-reported actions, and we provide an intuitive spatial map to visualise this association;

(4) feeling tones of not okay, boundary loss, annoyance or anxiety about tracking predict whether individuals take protective actions or not;

(5) more women feel negatively about tracking and report to not take any protective actions, compared to men, whose reports show that they are twice more likely to act protectively than women;

(6) less UK individuals expressed negative feelings and reported protective actions, compared to German and French individuals. UK individuals were also twice less likely to take any protective actions.


\emph{Outline.} In the rest of the paper, we review relevant literature, present the user study with the method and results, discuss the implications of our findings and conclude.

\begin{RedundantContent}
In addition to all Internet users potentially facing threats to their privacy, 
specific gender such as women, trans, gender diverse individuals may be more vulnerable.
I
While women tend to be more sensitive and concerned about their privacy than men, they are less involved in taking protective~\cite{oomen2008privacy,buchi2017caring} and technical actions~\cite{park2015men}. 
A gender gap in protective privacy behaviour is therefore likely prevalent online. 
\begin{RedundantContent}
It is also thought that general internet skills is an essential starting point for users' own actions 
towards protective privacy behaviour~\cite{buchi2017caring}, together with self-efficacy in (protective) skills~\cite{boerman2018exploring,wohn2015factors}. 
Although there are signs of narrowing of the gap in digital skills across gender, there are still some disparities online (such as across Europe~\cite{martinez2017digital}) and the gap in confidence remain wide~\cite{liberatore2020gender}. 
A gender gap in protective privacy behaviour is therefore likely prevalent online. 


This paper investigates two user-centred privacy problems prevalent online: (1) protection behaviour in relation to user experiences of tracking and (2) the gender gap in protective behaviour.
First, while previous research did not find a clear link between user understanding of tracking and their use of PETs~\cite{mathur2018characterizing},
we propose instead that users' tracking protection actions are associated their affective response.
We ground our proposition in literature on the informative and decisive value of emotions on behaviour~\cite{peters2006affect,vohs2007emotions,schwarz1990feelings}. 
We therefore investigate ``\emph{how do Internet users' feelings about tracking relate to their use of protective actions?}" 
Second, the availability of recent research on identifying the perception and practice of online privacy across gender is limited, and the underlying reasons which may contribute to protective differences across gender require in-depth investigation. 
We posit distinctions in affective response to tracking, protective actions, awareness and in use and perception of use of PETs across demographics of gender and therefore also investigate ``\emph{how do men and women differ in their feelings about tracking, and their protective actions, awareness and use of PETs?}"  


\emph{Contributions.} 
This paper contributes a relatively large-scale, gender-based and cross-national investigation of user tracking protection, to the rich user-centric privacy behaviour and tracking protection research area. 
It provides a two-pronged approach in eliciting individuals' protective actions, (1) via the study participants' free-form response, as well as (2) their responses to an exhaustive list of PETs.

\emph{First}, we find a significant association between feelings about third-party tracking (TPT) and self-reported actions, and 
provide a spatial map to visualise this association. 

\emph{Second}, we observe a privacy gender gap in protective behaviour, where women are more concerned, yet less likely to take actions, and men more likely to engage with more technical PETs. 
In addition, men prefer to use a self-supportive approach to protective behaviour, such as finding PETs 
via their own research, while women prefer a social- or collective-support approach, such as finding PETs with the help of family and friends.
Furthermore, women have weaker perception of use of PETs and perceived competency. 

\emph{Third}, we identify different groups of PETs based on their usage popularity and how individuals learn about them via a cluster map.
The visualisation groups $45$ PETs, into groups such as generic PETs that are in-effective for TPT, highly popular tracking-protection PETs that participants do not know how they learn about, or least popular PETs that are mostly found via research.
\end{RedundantContent}

\section{Background}
\label{sec:background}
We review literature on user attitudes, perceptions and protection methods with regards to tracking.
We then present research supporting the affective aspects of privacy, and review theories and research on behavioural responses to emotions. We complete this section with gender and national culture impacts on privacy.

\subsection{Tracking Attitudes, Understanding \& Protective Behaviour}
\label{sec:background_attitudes}
Individuals perceive behavioural advertising to be privacy-invasive~\cite{ur2012smart, mcdonald2010beliefs,turow2009americans}.
In particular, individuals
(1) do not want third parties to track and profile them online~\cite{rao2015they}, 
(2) are particularly concerned about the amount of data, the presence of sensitive information, and the data from offline sources found in tracked profiles~\cite{rao2015they},
(3) are sensitive to embarassing ads~\cite{agarwal2013not},
(4) are concerned about the lack of transparency and control over behavioural advertising practices~\cite{ur2012smart},
(5) are concerned that tracking could possibly lead to disadvantages in real life, and do not trust tracking~\cite{thode2015would}.
However, individuals also perceive behavioural advertising to be useful in providing ads that match their interests~\cite{ur2012smart,mcdonald2010beliefs,auxier2019americans},
enjoy its informativeness and utility for making purchasing decisions~\cite{schlosser1999survey},
feel comfortable in specific situations~\cite{melicher2016not},
and show varying acceptance of behavioural advertising depending on the context~\cite{chanchary2015user,wang2016examining,melicher2016not}.
In addition, research shows that individuals have basic understanding of online tracking~\cite{mathur2018characterizing} and
inaccurate and incomplete mental models of the mechanics of behavioural advertising, such as conceptualising trackers as viruses that access local files and reside on the local computer~\cite{yao2017folk}, and
having misconceptions about the purpose of cookies~\cite{mcdonald2010beliefs}.
With regards to protective behaviour, users want more control over tracking and perceive the benefits of controlled tracking, but are unwilling to put in the effort to control tracking, distrust existing tools~\cite{melicher2016not} or have limited awareness of the of countermeasures and how to use them~\cite{shirazi2014deters}, while some have reported to protect from tracking~\cite{pugliese2020long}.

\begin{RedundantContent}
\emph{Specific Protection Methods:} 
Among the protection strategies for behavioural advertising and tracking online, as categorised via the protection principle they rely on, by Estrada-Jimenez et al.~\cite{estrada2017online}, transparency and blocking are the most researched strategies in user studies.
Transparency refers to enhancing users' awareness of the tracking of their activities and data, such as via  \emph{MyAdChoices}, often accessible via the privacy policy or the cookie consent notice.
Blocking refers to limiting undesired interactions with third-parties and inhibiting known tracking mechanisms, and therefore advertising, such as via blockers as \emph{Adblock Plus} or \emph{Ghostery}.

Privacy notices of different sets of websites have been studied in the last few years \cite{Mehrnezhad,ValuePRivacy,utz2019informed,AsiaCCS,nouwens2020dark,matte2020cookie}.
It has been shown that the design of cookie notices (such as positioning, number of choices) substantially impacts whether and how individuals interact with consent notices~\cite{utz2019informed, nouwens2020dark}. 

Browser extensions vary in effectiveness and can be distinguished between (1) ad blocking extensions that limit ads from being loaded such, as \emph{Adblock Plus}, and (2) tracker blocking extensions that focus on blocking trackers, such as \emph{Ghostery}, \emph{Privacy Badger} or \emph{Disconnect}~\cite{merzdovnik2017block}.
In default settings, these extensions may not effectively block ads and trackers~\cite{pujol2015annoyed} and may need manual configuration for effective protection~\cite{wills2016ad}.
Although some extensions have improved their usability, their description were in the past found to be filled with
jargon and were not easy for users to change their settings when the tool interfered with websites~\cite{leon2012johnny}.
In addition, studies found that users download extensions for better user experience (UX) while possessing limited understanding of tracking~\cite{mathur2018characterizing} and because the extensions provided limited insight, users remained uncertain who the tracking companies are, what data they collect and for what purpose~\cite{schaub2016watching}.
\end{RedundantContent}

\subsection{The Affective Aspects of Privacy}
Although emotions are an integral aspect of decision-making and behavior,
research into emotional context of privacy is relatively new, and 
privacy decision-making has mainly been treated as a cognitive process. 
A limited number of research have demonstrated and argued that the emotional aspect of privacy is as important~\cite{farahmand2019privacy,stark2016emotional}, including
(1) FMRI neurobiological research providing evidence that privacy decisions are found to be in the more affective-cognitive area of the human brain than purely cognitive~\cite{farahmand2019privacy}; 
(2) research in the affective dimension of privacy attitude~\cite{coopamootoo2017whyprivacy};
(3) privacy decision-making research on the role of affect in disclosing personal information~\cite{li2011role}, in subconsciously shaping privacy risk perceptions~\cite{kehr2015blissfully} or overriding rational factors~\cite{knijnenburg2013making}; 
(4) research on the influence of discrete emotions (such as anger, anxiety, fear, regret) in leading to problem- and emotion-focused privacy coping strategies~\cite{cho2020privacy}; as well as 
(5) HCI research on the influence of emotional valence in interfaces on privacy concerns~\cite{kitkowska2020enhancing}.
These are supported by arguments from scholars who advocate for a greater attention to the phenomenology of feeling and to the concept of ``visceral" design in information privacy scholarship, policy, and design practice~\cite{stark2016emotional}.

\subsection{Behavioural Response to Emotions} 
\label{sec:Behavioural_Response_to_Emotions}

It is universally agreed that affect influences individuals' decisions, where human beings go through both cognitive and affective responses to stimuli. While cognitive responses indicate individuals' mental process in interaction with the stimuli, their affective responses designate individuals' emotional feedback from environmental cues.
Several models such as the affect-as-information hypothesis~\cite{clore2001affect}, the risk-as-feelings hypothesis~\cite{loewenstein2001risk}, or the use of affective heuristics when judging risky situations~\cite{slovic2002rational}, all explain the influence of affect in decisions and on behaviour. 

Affective evaluations are thought to reflect and translate the cognition of (privacy) concern or risk, leading to \emph{coping behaviour}~\cite{bechara1997deciding}, where coping is defined as an individual's efforts to manage stressful, aversive or disruptive events~\cite{lazarus1984stress}.
This is inline with the protection motivation theory that describes how individuals are motivated to react in self-protective ways towards a perceived threat and to adopt coping strategies~\cite{witte1992putting}. 

The couple of research into users' responses to affective privacy evaluations have 
investigated emotions associated with privacy risks and their resulting emotional coping (such as acceptance, avoidance, disengagement, or venting~\cite{cho2020privacy,jung2018investigation}), active problem-solving (such as company complaints~\cite{jung2018investigation}, negative-word-of-mouth~\cite{park2020users}, decreased usage time~\cite{park2020users}) or protective actions (such as engaging with privacy settings~\cite{jung2018investigation}), in contexts such as location-based services~\cite{jung2018investigation}, social networks~\cite{cho2020privacy}, and smart speakers~\cite{park2020users}.
They focused on negative emotions~\cite{park2020users,jung2018investigation,cho2020privacy}, such as anger, frustration, disappointment, anxiety, fear and regret, 
with the assumption that, as privacy loss and privacy risks contexts are generally considered as aversive rather than appetitive (or approach), they are more likely to elicit negative reactions 
than positive reactions~\cite{coopamootoo2017whyprivacy}.


\begin{RedundantContent}
On the one hand, FMRI neurobiological research provides evidence of human brain reactions to privacy risks, where privacy decisions were found to be more affective-cognitive than purely cognitive, supporting a dual-system process~\cite{farahmand2019privacy}.
On the other hand, scholars such as Stark, advocates for a greater attention to the phenomenology of feeling and to the concept of ``visceral" design in information privacy scholarship, policy, and design practice~\cite{stark2016emotional}.
Li et al.~\cite{li2011role} investigates role of both affect and cognition on online consumers' decision to disclose personal information.

Attitudes, including privacy attitude, have an emotional, cognitive and behavioral dimension~\cite{rosenberg1960cognitive,coopamootoo2017whyprivacy}. Compared to sharing attitude, privacy attitude is more associated with individuals' view of others as individuals (such as hackers) or organisations that can impeach on one's privacy, as well as a feeling of fear when evaluating one's privacy~\cite{coopamootoo2017whyprivacy}.
The effects of emotion in the privacy calculus is important, as privacy assessment is influenced by momentary
affective states, indicating that consumers often underestimate the risks of information disclosure
when confronted with a user interface that elicits positive affect~\cite{kehr2015blissfully}.
For instance, Kehr et al.~\cite{kehr2015blissfully} demonstrated that feelings and emotion subconsciously shape privacy-related risk perceptions and
that affect-based thinking is capable to not only shape but even override rational factors.
Knijnenburg et al.~\cite{knijnenburg2013making} also show how affect shapes privacy risk perceptions

    Recent research by Kitkowska et al.~\cite{kitkowska2020enhancing} into the effects of emotional valence (positive or negative) in privacy notices found that valence moderates the relationship between trust and privacy concerns, thereby further impacting information disclosure.
    They found that lower trust results in higher privacy concerns.
    However, affective state may alter the direction of this relationship. An increase in valence (more positive emotion) diminishes the effects of trust on privacy concern, which implies the possibility to alter privacy concerns through elicitation of emotions.
    Therefore UI with malicious intentions, such as with dark patterns with impact on emotional valence, could manipulate users' towards increased information disclosure.
\end{RedundantContent}


\subsection{Demographic Influence of Privacy} 
\label{sec:background_gender}
\begin{RedundantContent}
~\cite{epic2020gender,privacyinternational2020gender, weinberger2017sex},
where women are thought to be one of the most vulnerable demographics.
This phenomenon has been evidenced by (1) the privacy risks posed by women-targeted technologies that access intimate personal information that can then be used for monetary gain~\cite{MaryamCHI,shipp2020private}, as exemplified by menstruation apps which shared data about moods, cycle, sex life with Facebook and other third parties~\cite{privacyint2019nobodys};
(2) privacy-invasive technologies used to disproportionately or entirely target and objectify women, as seen by the presence of spy cameras or covert surveillance in areas where individuals expect privacy~\cite{airbnb2019hiddencamera,korea2018spycam}, facilitating voyeurism~\cite{hiddencamera2020uk} or spy camera porn, where $80\%$ of victims were women~\cite{korea2018spycam}; and 
(3) cyber-stalking and online harassment, where women were twice as likely as men to be targeted~\cite{duggan2017online}.
\end{RedundantContent}

Women tend to be more concerned about their general privacy~\cite{westin1997privacy,baruh2017online,sheehan1999investigation,youn2009determinants}, 
as well as privacy with regards to
specific technology (such as mobile devices~\cite{rowan2014observed} and online advertising~\cite{sheehan1999investigation}), compared to men.
However, there are nuances in response to concerns for men versus women, with regards to protective behaviour. 
Women have been found to be less confident and less equipped with technical skills to manage personal data~\cite{park2015men}, and more likely to use avoidance behaviours such as limiting the sharing of sensitive information~\cite{redmiles2018net}, while men have been found to be more familiar with the privacy protection strategies and to use them more often~\cite{oomen2008privacy}.
However women are thought to be as apt as men in social privacy protection strategies~\cite{park2015men,youn2009determinants}, such as taking protective actions in social network sites~\cite{blank2014new} or 
refraining from using websites that ask for personal information~\cite{youn2009determinants}. 

This distinction in concern and protective behaviour between women and men brings into focus the feminist perspectives of privacy 
~\cite{mackinnon1989toward,allen1988uneasy}, where women are thought to not enjoy the same level and types of desirable privacy online as men.
The conceptions of public/private spheres are thought to render women vulnerable, where they may have too much of the `wrong' kind of privacy, where a rigid binary classification of women versus men and surveillance also perpetuate patriarchal systems at the detriment of women, who are significantly identified and revealed but do not have agency in expressing and exercising their privacy.  


National culture, as the collective mindset distinguishing members of one nation from another~\cite{hofstede1984culture}, also influences privacy concern, behaviour or valuation of information~\cite{cho2009multinational,reed2016thumbs,thomson2015socio}.
Between the UK and Europe, there are indications of the British exhibiting different privacy technology usage behaviour compared to other countries, where in contrast, German users are thought to be better versed with `advanced' privacy technologies~\cite{coopamootoo2020usage}.
Compared to other countries (US or EU), German nationals attribute a higher importance to data protection~\cite{soffer2014privacy}, in sensitivity of data collected~\cite{cvrcek2006study} or in the perceived risks of privacy violations~\cite{krasnova2010privacy}.

\section{Related Research \& Gaps}
\label{sec:related_research_and_gaps}
In this section we compare our contributions to the most closely related previous research.

\emph{User-centred aspects of tracking:}
Previous research has mostly inquired into the cognitive dimension of tracking, such as via user perceptions~\cite{mcdonald2010beliefs, turow2009americans, ur2012smart,yao2017folk}, 
attitudes~\cite{schlosser1999survey} 
and concerns~\cite{rao2015they,wohn2015factors} 
of online behavioural advertising for insights into tracking.
A few investigations have focused their elicitation methods particularly towards tracking perception and mental model~\cite{shirazi2014deters,mathur2018characterizing,chanchary2015user}, 
concerns~\cite{agarwal2013not}, 
preferences~\cite{melicher2016not}, 
or third-party tracking online~\cite{thode2015would}. 

While feelings with regards to tracking have come up in investigations of perceptions and concerns about behavioural advertising and tracking (such as `creepy' or `scary')~\cite{ur2012smart}, to our knowledge, previous research has not explicitly asked individuals how they \emph{felt} about third-party tracking, that is, with an intention to specifically elicit feeling tones.
This current research explicitly investigates feeling tones with an intention to map these with protective actions.

For protective behaviour related to tracking, research has qualitatively elicited protective actions and the use of privacy technologies in general~\cite{shirazi2014deters,chanchary2015user,pugliese2020long}, or queried use of specific technologies such as browser extensions~\cite{mathur2018characterizing}. 
We use an open-ended method to elicit individuals' protective actions.

\emph{Behavioural Response to Affective Privacy Evaluation:} 
Previous studies~\cite{jung2018investigation,cho2020privacy} have focused on technology contexts that are different to this paper, as reviewed in Section~\ref{sec:Behavioural_Response_to_Emotions}.
In addition, these studies query participants to a preset list of discrete negative emotions and coping behaviours, whereas we extract feeling tones and protective actions from participants' free form response.

\begin{RedundantContent}
\section{Research Aim}
\label{sec:aim_feeling_action}
We conduct a relatively large-scale user study online.
In this section, we provide the motivations and research questions, as well as describe the overlaps and the additions we make to the research field.

While previous research has \emph{not} found a link between individuals' understanding (or mental models) of tracking and their use of protective technology (such as the use of browser extensions)~\cite{mathur2018characterizing}, we posit that \emph{feeling tones} can better predict protective actions.
We focus on feeling tones, rather than traditional measures of emotions or affect~\cite{watson1999panas},
because we aim to investigate the mood or feeling associated with the particular experience (or stimulus) of TPT (inline with APA dictionary's definition of feeling/affective tone~\cite{APA-Dictionary}).

We support our proposition in (1) research on the influence of affect in decision-making outcomes~\cite{peters2006affect,vohs2007emotions},
(2) research on the informative value of feelings, where the feeling-as-information theory conceptualises the role of experiences in human judgment~\cite{schwarz1990feelings}, and 
(3) on the fact that attitudes are multidimensional concepts with components of cognition, emotion and behaviour~\cite{fazio1986attitudes}.
In particular, privacy attitudes are thought to comprise feelings of fear (anxiety) and anger (annoyance)~\cite{coopamootoo2017whyprivacy}.

In addition, individuals across countries may use different privacy protection practices~\cite{coopamootoo2020usage} and have different experiences of third-party tracking, where in particular for individuals from the UK, Germany and France,
there may be variation in behaviour due to differences in the importance attributed to data protection~\cite{soffer2014privacy}, in sensitivity with regards to the duration or quantity of data collected~\cite{cvrcek2006study} or in the perceived risks of privacy violations~\cite{krasnova2010privacy}.

Furthermore, men and women (and non binary gender) may exhibit different experiences of tracking and privacy protection behaviour~\cite{park2015men}, where previous research point to men being more familiar with various privacy protection methods and user them more often than women~\cite{oomen2008privacy}.

We therefore investigate

\textbf{RQ1} ``\emph{How do individuals feel with regards to third-party tracking?}", given their gender and country differences.

\textbf{RQ2} ``\emph{What tracking protective actions do individuals employ online?}", given their gender and country differences.

\textbf{RQ3}: ``\emph{How are individuals' feelings about third-party tracking associated with their protective actions?}"

\textbf{RQ4}: ``\emph{How do individuals’ feelings predict whether they take protective actions or not, given their gender and country differences?}"
\end{RedundantContent}

\section{Method}
In this section we provide details about participant recruitment and characteristics, the study procedure and questionnaire design, the research ethics process, as well as the design limitations. 

\subsection{Participants}
We recruited participants via Prolific Academic, a crowd-sourcing platform whose
data quality has good reproducibility~\cite{peer2017beyond} is comparable to Amazon Mechanical Turk's which is widely used within security and privacy user studies.
The study lasted between $20$ to $30$ minutes. Participants were compensated at a rate of \pounds$7.5$ per hour, slightly above the minimum rate of \pounds$5$ per hour, as advised by Prolific Academic.

We sampled around $630$ participants and ended with $N=614$ after removing incomplete attempts at the survey.
The study was balanced by number of participants in each country and gender.
The $N=614$ participants consisted of $n=209$ from the United Kingdom (UK), $n=202$ from Germany (GE) and $n=203$ France (FR).
We chose these three countries as they have the highest number of internet users in Europe~\cite{statista2019internet} and therefore a high number of users potentially exposed to online tracking.
While nationals of different countries, such as the UK, Germany and France, may exhibit different privacy behaviour~\cite{coopamootoo2020usage}, 
we note that the UK has similar data protection provisions as the rest of Europe, as it has established its implementation of the GDPR, where the principles, rights and obligations of the `UK GDPR' follows the European one~\cite{parliament2018data}. 
Overall there was approximately a similar number of women ($n=307$) and men ($n=299$) participants (about $100$ each in each country) and $8$ self-reporting as non-binary.
The rationale for balancing across gender is that women may engage in different protection practices compared to men, who are known to be more familiar with protection methods~\cite{park2015men}. Women can be considered as a vulnerable user group, where as explained by the `differential vulnerabilities' concept that recognizes how different populations face different types and degrees of security and privacy risks~\cite{Pierce}, 
they may experience more harmful impact from the same online threats, compared to men. We posit that such user groups need dedicated research and support. 

Table~\ref{tab:demo_study1} provides a summary of the demographic details.
The table also lists the most used browsers, where
`other'  included mentions of DuckDuckGo, Tor, Microsoft Edge, Chromium or Vivaldi. 
\begin{table}[t]
    \centering
    \caption{Participant Characteristics}
    \label{tab:demo_study1}
    \footnotesize
    \begin{tabular}{lrcccc}
        \toprule
        \textbf{Country}         & \textbf{$N$}  & \textbf{Mean Age}  & \multicolumn{2}{c}{\textbf{Gender}}                                               \\
        \cline{4-6}
                                 &               &                    & \#F                            & \#M                      & \#N  \\
                         &&&&        &                                                                                                                       \\
        United Kingdom       & 209           & 35.78              & 109                                 & 100                         & 0             \\
        Germany              & 202           & 29.21              & 100                                 & 100                         & 2             \\
        France               & 203           & 27.29              & 98                                  & 99                          & 6             \\
        \midrule
        \textbf{Education} & \textbf{$\%$} & \textbf{Ethnicity} & \textbf{$\%$}                       & \textbf{Most Used } & \textbf{$\%$} \\
          &&&&   \textbf{Browsers}                    &                                                                                                                        \\
        High School/lower     & 26.1          & white              & 86.6                                & IE           & 10.4          \\
        College                  & 17.6          & mixed              & 5.0                                 & Chrome                      & 72.3          \\
        Undergraduate      & 29.0          & asian              & 4.1                                 & Firefox                     & 34.0          \\
        Masters            & 24.9          & black              & 2.1                                 & Safari                      & 19.4          \\
        Phd                      & 2.2           & other              & 2.1                                 & Opera                       & 5.7           \\
                                 &               &                    &                                     & Brave                       & 9.4           \\
                                 &               &                    &                                     & Other                       & 6.7           \\
        \bottomrule
    \end{tabular}
\end{table}

\subsection{Procedure} 
We ran the study as an online survey during 2020. Participants were first presented with (1) a consent form (as described in Section~\ref{sec:ethics} below), (2) followed with a demographics questionnaire,
(3) elicitation of their feelings with respect to tracking online, and 
(4) their own protective actions.
The survey 
was proof-read by the authors, 
and $3$ of their acquaintances who are not experts in the topic. In addition, the survey was piloted on Prolific Academic across the three countries, where we invited $9$ pilot participants, with 95\% approval rate, to comment on the survey, thereby facilitating enhancements.

The first page of the survey gave information about the study, letting participants know that the study was anonymous, that participation was voluntary, and explicitly asked for opt-in consent for participation.

Next, we describe the feelings and protective actions elicitation, as well as provide the questions verbatim. 
These were set as open-ended questions with participants responding in free-form text.
\emph{\textbf{Feelings:}}
\label{sec:questionnaire_experience-action}
To bring participants' own experience and feelings about tracking to the fore, we used a method similar to a mood induction protocol~\cite{westermann1996relative} to elicit feelings about an issue.
We first asked participants to write about their understanding of third-party tracking as a way of inducing their mental picture of third-party tracking.
We then asked them to express their feelings about third-party tracking on the web in their words, in writing.
The two questions were set as: 
(1) ``\emph{In your own words, write about your own understanding of third-party tracking on the web. 
In particular, what does third-party tracking mean?"}
followed with (2) ``\emph{How do you feel with regards to third-party tracking on the web.
Please name emotions and/or perceptions if relevant.
With regards to third-party tracking, I feel ... }".
Note that we focus on feeling tones, rather than traditional measures of emotions or affect~\cite{watson1999panas},
because we aim to investigate the mood or feeling associated with the particular experience (or stimulus) of tracking (inline with APA dictionary's definition of feeling/affective tone~\cite{APA-Dictionary}).

\emph{\textbf{Protective Actions:}}
We followed with a question to elicit participants' tracking protection actions, where we asked
them to name the actions they employ. In particular, we asked ``\emph{What actions have you taken to protect yourself from tracking (including third-party tracking), as you browse the web?}".

After data collection, incomplete attempts at the survey, making up $16$ participants, were removed leaving us with $N = 614$ participants.
We then progressed into qualitative analysis of the free-form responses to the feelings and actions questions. 
We provide description of this analysis, including the creation of a codebook, in Section~\ref{sec:qual_coding}.
We used these codes to report on feeling tones and actions, across gender and country, as well as within the quantitative analyses that follow.
We summarise the study design in Figure~\ref{fig:study_design}.

\begin{figure}[h]
\centering
\includegraphics[keepaspectratio,width=.99\columnwidth]{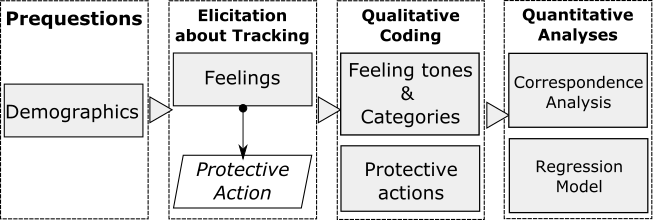}
\caption{Study design}
\label{fig:study_design}
\vspace{-.65cm}
\end{figure}


\subsection{Ethics}
\label{sec:ethics}
We obtained full approval from Newcastle University's Ethics Committee before the research commenced. We also sought participants' opt-in consent for data collection prior to their responding to the questionnaire.
In addition to having undergone independent ethical review, we designed our user studies to address pillars of responsible research in computer science~\cite{bailey2012menlo}. 
Participation in the study was voluntary and anonymous and our participants could drop out of it at any stage. 

\subsection{Limitations}
We discuss the limitations of the study design as follows.

We queried protective actions in participants' own words, which assumes at least a minimum awareness and understanding of the actions they employ, rather than providing a pre-defined list of privacy technologies for participants to choose from.
However, we specifically queried participants on protective actions and did not capture other behaviours that they may employ for emotional coping, as described in Section~\ref{sec:Behavioural_Response_to_Emotions}.

The study relies mainly on self-reports, which is a widely used and valuable form of eliciting user responses in privacy and security user studies.
While self-report can be argued to induce bias, previous research has also found that self-report insights can translate to real-world settings~\cite{redmiles2018asking}. 
In addition, free-form responses provide a rich view of users' own experience.

By choosing to position the demographics questionnaire at the beginning of the survey, the study may have been affected by `stereotype threats'~\cite{hughes2016rethinking}. We would therefore opt for a different placement of demographic questions in the future.

Although the study was piloted across the countries, it was written in English, and therefore has minor limitations on who can take the survey.
Future studies targeting different nationals across countries may consider using their first language. 

Because the number of non-binary participants was only $8$, our gender comparisons and statistical analyses focus on women versus men. We hope to target a more diverse sample in future research.
In addition, while other user characteristics (such as skills) may affect experience and protective actions, we focused and balanced our sample on gender and country demographics. A future study may compare the influence of a further list of user characteristics.
Furthermore, although having a relatively large-scale qualitative grounding, the study was directed at the UK and European countries, thereby limiting generalisability to other national cultures.



\section{Qualitative Coding}
\label{sec:qual_coding}
In this section, we describe our process of extracting participants' feeling tones and actions from their responses.

\subsection{Feeling Tones}
\emph{Identification of Themes \& Codebook Creation.}
We looked into participants' free-form text responses.
We used a conventional line-by-line coding method (as previously employed in user-centric privacy research~\cite{coopamootoo2017whyprivacy}), where we read each response and identified specific themes.
We know from previous research that individuals find tracking to be worrisome (scary~\cite{ur2012smart}), embarassing~\cite{agarwal2013not}, have mixed feelings about tracking~\cite{chanchary2015user,wang2016examining,melicher2016not} or are okay with tracking~\cite{ur2012smart,mcdonald2010beliefs}.
We therefore sought to identify whether participants were in general, (1) okay or do not care about tracking, (2) not-okay or feel negatively (3) sometimes okay, sometimes not-okay or okay under certain conditions. 
We further looked into the emotions and/or reasoning expressed by participants, and also identified `other' themes that were not categorised in the three categories.
We note that while some participants only expressed how they felt about tracking, most participants provided a reasoning in association to their feelings. We detail these in the results Section~\ref{sec:feelings}.
We summarise the feeling tones according to their categories, as well as the \% participant responses that express them, in Table~\ref{tab:category_feelings}. 
We also provide the whole codebook with example words in Appendix~\ref{sec:ap_codebook} in Table~\ref{tab:feelings_codebook}.

\begin{table}[h]
\centering
\caption{Categories and Example Tones (N=614)}
\label{tab:category_feelings}
\footnotesize
\begin{tabular}{l|rr}
\toprule
\textbf{Tone Category} & \textbf{Feeling Tone} & \textbf{\% Participants} \\ \midrule

& Not\_Okay &4.6 \\
 & Boundary\_loss & 17.8 \\
Generally & Unfair &8.6 \\
not okay& Annoyance &12.5 \\
(Negative) & Anxiety &12.4 \\
& Discomfort &9.1 \\
& Distrust & 3.9\\ \midrule

Sometimes okay, & Ambivalent & 7.7\\
sometimes not okay & Okay\_if & 2.9\\
 & Okay\_ protected & 2.3\\ \midrule

Generally & Okay & 4.7\\
okay /  & Indifferent & 2.9  \\
indifferent & Necessity & 0.7  \\
& None & 1.6\\ \midrule

Other & Should\_regulate & 2.9 \\
& Not\_aware & 5.4 \\
\bottomrule
\end{tabular}
\end{table}

\emph{Coder Reliability.}
One researcher created the codebook, which was then provided to a second researcher along with a sample of responses.
The codebook was iteratively refined, after which two researchers coded the whole set of responses.
We computed $95\%$ agreement between coders as well as Cohen $k$ of $.820$, $p<.001$, suggesting substantial agreement in applying the codebook.

\subsection{Protective Actions}
We used a line-by-line coding to extract words such as `browser option', `extension' or specific names for privacy-enhancing technology (PET) such as `Ublock', and categorised these into tracking protection methods. Because the coding involved simply identifying specific words relating to tracking protection, it was conducted by one researcher only. 
Participants also mentioned strategies that we grouped under `\textbf{other}'. These included actions such as `checking social media settings or avoiding public Wi-Fi. We describe these in Section~\ref{sec:protective_actions_self_report}.
We summarise the protective actions by \% participants naming them in Table~\ref{tab:actions}.
Note that some participants named more than one protective action.

\begin{table}[h]
\centering
\caption{Protective Actions (N=614)}
\label{tab:actions}
\footnotesize
\begin{tabular}{r|r}
\toprule
\textbf{Actions} & \textbf{\% Participants} \\ \hline
extension	& 27.7 \\ 
clear cookies&	16.6\\ 
private browsing&13.5 \\
VPN	&11.1 \\
builtin browser setting & 9.6\\
clear browser history& 3.3\\
anti-malware, -virus	&2.6\\
safe website&	2.1\\
other& 8.0 \\
&\\
No Action	 &34.7\\ 
\bottomrule
\end{tabular}
\end{table}

\section{Results}
We report our qualitative and quantitative findings.
We note that for gender-based comparisons, to aid statistical reporting, we focus on male versus women only, as only $8$ out of $614$ participants self-reported as non-binary.

\subsection{Feelings about Third-Party Tracking}
\label{sec:feelings}
We investigate RQ1, that is ``How do individuals feel with regards to third-party tracking?", given their gender and country differences.
We summarised the feeling tones extracted from participants' responses and their categories in Table~\ref{tab:category_feelings}.
In the following subsections, we (1) provide the \% of participants expressing the feelings in each category, across gender and country and then we (2) describe the extracted feeling tones, while providing example responses from participants.
In the example responses, we refer to UK participants as UK\#, German participants as GE\# and French participants as FR\#.

\subsubsection{Generally Not Okay (Negative Feelings)}
\begin{figure*}
	\centering
	\includegraphics[keepaspectratio, width=1\textwidth]{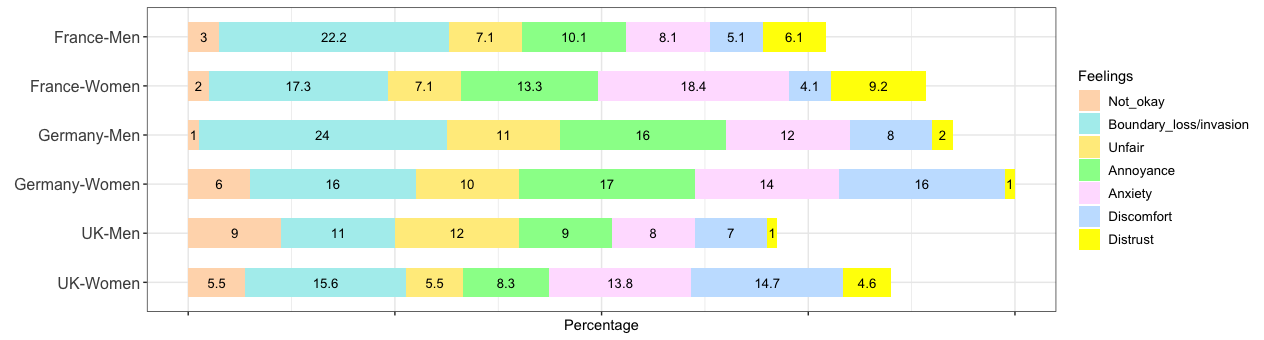} 
	\caption{\emph{Negative} feelings expressed across gender \& countries 
	(the x-axis shows \% in each gender-country group with n=100 approx.)
	\footnotesize{\emph{See text for description of noticeable patterns, such as more women feel negatively \& less UK participants feel negatively.} 
	}}
 \label{fig:negative_feelings}
 \end{figure*}

Participants' whose feeling tones were in congruence with not being okay with third-party tracking (referred to as TPT in the results section) either just said so, named (their concern with respect to) the threats of TPT (such as invaded privacy or self/boundary loss, unfair practice by others) or went further to describe their emotional response as a result of TPT (such as anger, anxiety, distrust or discomfort). 
 
Figure~\ref{fig:negative_feelings} summarises the feelings tones categorised under generally not okay or negative feelings.
We notice patterns across gender and country, for example that (1) more women expressed negative feelings in all countries; 
(2) less UK participants expressed negative feelings compared to Germany and France, across both gender; and
(3) slightly more men expressed the mix of feelings till `annoyance', but more women expressed `anxiety, discomfort, distrust' (combined).

\textbf{Not Okay:}
$4.6\%$ of participants said they were not okay with TPT, 
such as by saying that they did not feel good about TPT (without mentioning particular emotions), or clearly stating that they do not like or want TPT and are against TPT.
Example responses include:
GE179 ``\emph{\textbf{I do not feel very good about it}, because I don't know where my data will end up}",
UK20 ``\emph{\textbf{I don’t like it} and I don’t like other companies having my information}",
FR116 ``\emph{If they're what i think they are, then \textbf{i really don't like this business model} [sic]}",
UK114 ``\emph{\textbf{I dislike that you have to actively log out of offering this access to third parties}...}", 
FR113 ``\emph{\textbf{opposed}}",
UK164 ``\emph{\textbf{I am strongly against} unannounced third party tracking...}", and
FR128 ``\emph{\textbf{I would rather not be tracked at all}}".

\textbf{Boundary loss and invasion:} $17.8\%$ participants described feeling an invasion or violation with regards to their online privacy, such as expressed by
UK36, ``\emph{That it is an \textbf{invasion of your personal details} and that keeping it to just those companies that you choose to deal with would be ideal but you know that info will probably end up being passed on...}",
by UK85 ``\emph{this is a \textbf{violation of privacy and of freedom}}", or
UK90 ``\emph{An invasion of privacy, masqueraded by third party/flimsy data protection rules}".

This feeling tone also included participants who went further to express how they felt with regards to the loss of boundaries, in particular noting the consequences to their personal sphere or sense of self, such as expressed by
GE1 ``\emph{With regards to third-party tracking, I feel \textbf{exposed} to the web}",
UK68 ``\emph{slightly out in the open}",
UK51 ``\emph{I feel \textbf{violated} and as though my privacy and security is not respected}".
GE25 ``\emph{\textbf{unsecure} since I value my privacy. I don't want third partys [sic] to get access to any of my personal information},
GE27 ``\emph{exposed, \textbf{exploited}}",
GE66 ``\emph{\textbf{unsafe}, spied on}",
GE134 ``\emph{I feel like my every move is being \textbf{monitored}}", and
UK115 ``\emph{I don't like the idea of being \textbf{stalked} online, however that may be}".

Additionally, some participants from Germany and France used even stronger wordings to describe feeling deeply invaded as a person, or dis-humanised, such as
GE7 ``\emph{unsafe. I feel \textbf{naked.} `transparent human' we call that in German [sic]. I feel like nothing is private anymore, even if I seek the anonymity of the internet for a reason}",
FR4  ``\emph{I feel \textbf{raped}, robbed, angry}",
FR143 ``\emph{I feel this is very \textbf{intrusive} and it is not very moral}",
FR82 ``\emph{vulnerable}",
GE135 ``\emph{Unsafe and somewhat \textbf{inhuman}, it feels like I'm treated as just another customer for product XY}",
GE139 ``\emph{exploited, unsafe, \textbf{like an object}}".

\textbf{Unfair practices:} $8.6\%$ of the participants focused on the methods practised by websites or companies that result in unfair or helpless situations such as expressed by
UK1 ``\emph{there should be less third party tracking because sometimes \textbf{it is not you making a decision}, it is the adverts telling you to make a certain decision}",
UK3 ``\emph{baddies, trying to \textbf{steal my info for their own illicit purposes}}",
UK8 ``\emph{\textbf{betrayed} by the companies that sell my data}",
UK25 ``\emph{... it feels a little \textbf{dishonest/sneaky}}",
GE46 ``\emph{surveilled and powerless. I have \textbf{no opportunity to disagree} to the tracking except not using the website...}", 
GE55 ``\emph{\textbf{fucked because they can do whatever they want} and nobody stops them}",
GE109 ``\emph{..\textbf{that I do not have privacy on the internet that I want} or that I am supposed to believe I have. I do not want companies to track my activities on the internet or sell my data but I need to agree to do so in order to use some services on the net}", or
FR4 ``\emph{...Feel like we are \textbf{trapped}}".

When participants went beyond expressing privacy/self invasion or unfair practices to name their emotional response to their dismay with TPT, we coded these under the specific emotion named, such as annoyance, anxiety, discomfort or distrust. 

\textbf{Annoyance:} $12.5\%$ participants named feeling tones traditionally categorised under anger-related emotions~\cite{watson1999panas}, such as annoyance, irritation, disgust or exasperation. These were expressed as emotional response to unfair practices, privacy invasion, lack of transparency, due to the presence of ads, or because something better should be offered.
Example responses include:
UK10 ``\emph{\textbf{irritated}, unsecure, harassed, \textbf{annoyed, not happy} for them using my information without permission}",
UK64 ``\emph{\textbf{annoyed by it, that I'm being spied on}}",
GE50 ``\emph{Bothered, annoyed, stalked, disrespected}",
GE44 ``\emph{Mostly I feel that third-party cookies are \textbf{annoying and not nearly transparent enough}, even if you're notified of them once you visit a website}", or
GE20 ``\emph{that it tends to get annoying. once you search something out of curiosity your ads might be spammed with this very product that you in reality dont [sic] have the biggest interest in}".

\textbf{Anxiety:} $12.4\%$ participants named feeling tones associated with anxiety, using words such as scary, worry, anxious, cautious, creepy or spooky.
These were expressed as emotional response to the amount of information about someone that can be left online,
the `not knowing' about what TPT exactly is, how it happens, who accesses personal data and how to protect, and also when signs of tracking are noticed online.
Example responses include:
UK18 ``\emph{That it is a \textbf{pretty scary thing} seeing how much of yourself you leave on the Internet...}", 
UK41 ``\emph{\textbf{Worried} that a third party that I don’t know about is accessing my information}",
UK53 ``\emph{It's pretty relevant in modern society with the increase in technology use, it's \textbf{frightening} to not know enough about it to fully protect myself}",
UK79 ``\emph{It would make me feel \textbf{cautious} about what information I am sharing online}",
UK107 ``\emph{It sometimes \textbf{creeps me out} when I browse the internet for say make-up products and then the same webistes [sic] I visit appear in adverts on my Twitter feed...}'',  
UK112 ``\emph{definitly [sic] makes me a bit nervous and a bit creepy almost like my phone is always listening to me}", or
UK142 ``\emph{...it is a bit \textbf{spooky} to find links to things that you have been searching for and I do not like it at all}".

\textbf{Discomfort:}
$9.1\%$ participants reported feeling discomfort as a result of privacy loss or the lack of choice, not feeling at ease as they are not fully aware of the impact, and uncomfortable at the signs (by ads) that they are tracked.
Participants used words such as uncomfortable, unsettled, overwhelmed, unpleasant, uneasy or disturbed. Example responses include:
UK28 ``\emph{\textbf{not comfortable}, i feel them to be intrusive. [sic] We don’t have a choice, if we want to use a site ...}", 
GE23 ``\emph{\textbf{Uneasy}, I don't want to have Facebook tracking me on pages other than their own}",
UK67 ``\emph{\textbf{unsettled by it} as I don't know who the third party is or what they are doing with my data}", and
FR84 ``\emph{I feel \textbf{rather uncomfortable} when I come across ads that are clearly oriented towards content that I've already searched for on the web}".

\textbf{Distrust:} $3.9\%$ participants expressed a lack of trust in websites or companies online, 
thereby expressing suspicion, wariness, or lack of trust.
Example responses include:
GE81 ``\emph{I feel it is a very \textbf{unknown and suspicious business}. Since you mostly agree via one click on a page long agreement that you haven't read carefully which results in unknow [sic] persons and institutions using your information}",
FR32 ``\emph{I feel really suspicious, I do not rely on cookies and block them as possible as I can}",
FR54 ``\emph{\textbf{distrustful}, attentive, but non-paranoid}", and
FR115 ``\emph{I am \textbf{wary about being tracked online}. I don't feel safe, I feel like I'm being exploited for my information}".

\subsubsection{Sometimes Okay, Sometimes Not Okay}
A group of participants were either sometimes okay and sometimes not-okay (ambivalent) or
okay under certain conditions (okay\_if) or okay because they were already protected (okay\_protected).
We visualise the differences between gender and country in Figure~\ref{fig:ambi_feelings} and notice the following patterns:
(1) more men than women expressed `ambivalence' or combined `ambivalence, okay\_if, okay\_protected' feelings in all countries; and (2) more UK participants expressed ambivalence compared to Germany and France, across both gender. 

\begin{figure}[h]
	\centering
	\includegraphics[keepaspectratio, width=1\columnwidth]{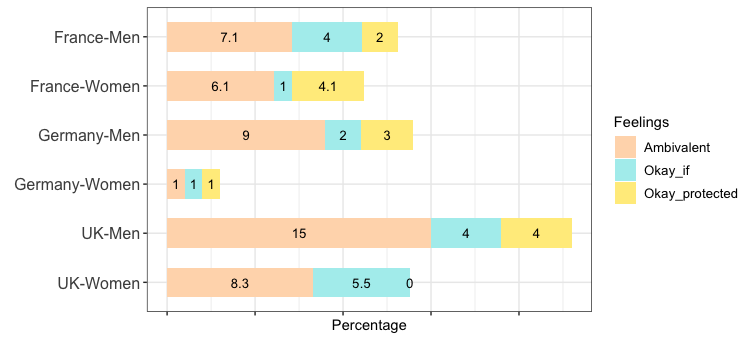} 
	\caption{\emph{ambivalent} or {okay under certain conditions} feelings (the x-axis shows \% in each gender-country group with n=100 approx.)}
	\label{fig:ambi_feelings}
\end{figure}

\textbf{Ambivalent}:
7.7\% participants expressed that TPT can be both positive and negative.
Example responses include:
UK6 ``\emph{\textbf{It can be okay sometimes but it’s a bit bad that it can happen}}",
UK24 ``\emph{Pritty brutal [sic], but its the only way people make money on the web with ads anymore, with old people clicking the ads}",
UK58 ``\emph{A little intruded on but accept its a part of how the internet works}",
GE26 ``\emph{exited [sic] at having an application that knows my desires better than i do, and a little scared at the exact same point}, and
GE97 ``\emph{Ambivalent because it enables more free services, but can also be intrusive on privacy}".

\textbf{Okay, if}: Another 2.9\% of participants expressed that they would be okay with TPT under certain conditions, such as personal data being used only to provide ads.
Example responses include:
UK13 ``\emph{...I don't mind seeing ads that are relevant to me and as long as the third party is not using the data in a harmful way then I don't mind}",
GE35 ``\emph{Okay about collection of user data for advertising purposes as long as it is purely commercial and not political}", or 
UK23 ``\emph{It's ok as long as its not intrusive and no data is kept}".

\textbf{Okay, protected}: 2.3\% participants expressed being okay with TPT because they took protective actions.
Example responses include:
FR177 ``\emph{I do not care because I use and adblocker [sic]}",
GE64 ``\emph{relaxed because I block the most of it using DNS based blocking and adblockers}", and
FR52 ``\emph{I use brave [sic] for this very reason but i know that sometimes you can access some website due to them blocking you if you don't accept the tracking}".

\subsubsection{Okay or Indifference} 
A group of participants were okay with TPT, were indifferent, thought TPT was a necessity, or said that they felt nothing.
We visualise the participant breakdown across gender and country in Figure~\ref{fig:okay_feelings} and notice that 
(1) slightly more women than men said that they had no feelings across all countries; and
(2) more men were okay or indifferent, except for Germany.

\begin{figure}[h]
	\centering
	\includegraphics[keepaspectratio, width=.99\columnwidth]{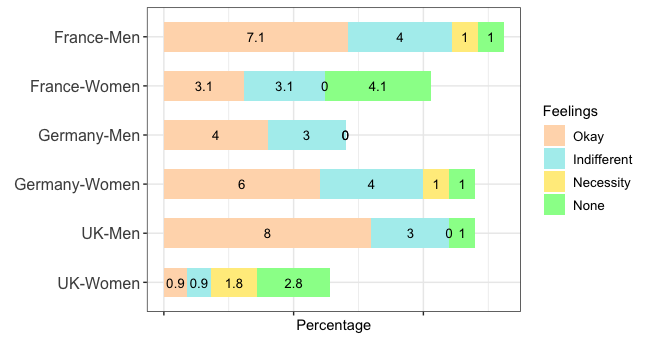} 
	\caption{\emph{okay or indifferent} feelings (the x-axis shows \% in each gender-country group with n=100 approx.)}
	\label{fig:okay_feelings}
\end{figure}

\textbf{Okay:}
$4.7\%$ of participants said they were okay with TPT, in particular expressing that TPT was alright, or expressed positive feelings such as being content and comfortable with TPT, or focused on the advantages of TPT such as personalised ads, and a lack of concern.
Example responses include: 
FR10 ``\emph{I understand that third party tracking is a part of what makes a lot of content on the internet free. \textbf{I feel ok with it}}",
UK2 ``\emph{\textbf{I feel its alright i have no bad feelings towards it}}",
GE13 ``\emph{\textbf{okay with it since it's only my online behavious that is being obersevd} [sic]. In addition most of the data collected is used for machine learning and not reviewed by actual humans}",
UK5 ``\emph{I don't mind it}",
FR88 ``\emph{\textbf{comfortable}, I have no problem with that}",
GE130 ``\emph{I sometimes like advertisments that \textbf{carter to my taste}}", 
GE16 ``\emph{\textbf{used too}, i think the web knows so much from me, i have nothing to hide, ok they can get this data too}", or
GE93 ``\emph{\textbf{not really concerned}, since those data might be valuable for marketing, but don't effect me personally in any way}".

\textbf{Indifferent, necessity, none:}
$2.9\%$ of participants said they were \textbf{indifferent} to TPT, such as by expressing indifference or not caring such as
FR64 ``\emph{I feel a bit indifferent to be honest}",
UK45 ``\emph{Not bothered}",
GE90 ``\emph{Although it may be morally debatable, nobody really seems to care about it}",
GE131 ``\emph{With regards to third-party tracking, I feel like i couldn't care less, honestly}".
$0.7\%$ participants spoke of the \textbf{necessity} of tracking online and of their acceptance of it, such as
UK49 `\emph{That it is just a necessary part of being online these days}",
UK119 `\emph{The amount of information out there is way too much to understand for a generic user. I accept that using the internet entails such addition to it}".
$1.6\%$ participants said that they felt nothing, such as 
UK65 `\emph{Have no feelings}",
GE124 `\emph{no particular positive or negative emotion}",
FR42 `\emph{I don't feel any particular way about it}".

\subsubsection{Other}
Another group of responses did not point to emotional evaluation of TPT.
Instead they called for regulation or said they were not aware of TPT. We grouped them under `other'.

\textbf{Call for Regulation:}
Although the GDPR makes provision for tracking protection, 
$2.9\%$ of participants felt a need for (stronger) regulation, for completely banning TPT, or for particular legal coverage (such as opt-in only or more transparency). These responses show Internet users' un-awareness vis-\`a-vis the GDPR's coverage or their dis-satisfaction that the existing regulation do not do enough for privacy.
Example responses included:
FR43 ``\emph{I think it's a good idea but \textbf{it has to be regulated}}",
GE196 ``\emph{very dissatisfied. \textbf{Should be forbidden!} [sic]}",
UK194 ``\emph{It \textbf{should be illegal}. I don't see how it can even be legal in the first place},
UK123 ``\emph{It is impossible to avoid as they cookies [sic] on the on every website which all communicate with each other this \textbf{should be banned}},
GE103 ``\emph{Like it \textbf{should be opt-in only}}",
FR40 ``\emph{that it \textbf{should be made transparent} who can track us and for what reason, with an opt out}", and
UK66 ``\emph{I think it is irresponsible and people \textbf{should be made aware very clearly} that they are being tracked, and can be damaging to people's mental health}".

\textbf{Not aware:}
About $5.4\%$ of participants said they were not aware of TPT or were confused by it, including
UK48 `\emph{No idea What this is [sic] I don’t really take any interest in being safe on the Internet I mean I probably should but I don’t}", or 
UK77 `\emph{I don't know enough about this}".

\begin{RedundantContent}
 \begin{figure*}[ht]
	\centering
	\includegraphics[keepaspectratio, width=.99\textwidth]{./figures/Rplot_emo_balloons_3countries} 
	\caption{Comparative view of full set of feelings across gender and countries}
 \label{fig:all_feelings1}
 \end{figure*}

\textbf{Demographic Comparison}
We investigate RQ2a, that is ``how do individuals in the UK, Germany and France differ in their experience of TPT?"
We summarise the feeling tones expressed by the $\%$ of participants in each country in Figure~\ref{fig:user_emotions}.
We compute a $\chi^2$ test and observe a significant difference between countries for feeling tones of annoyance ($p=.042$), discomfort ($p=.017$), distrust ($p=.007$), and ambivalence ($p=.032$).
For RQ2b, with regards to overall men versus women, we observe significant differences, where more women reported anxiety ($p=.018$) and discomfort ($p=.022$), while more men reported ambivalence ($p=.013$).
\end{RedundantContent}

 \begin{figure*}
	\centering
	\includegraphics[keepaspectratio, width=1\textwidth]{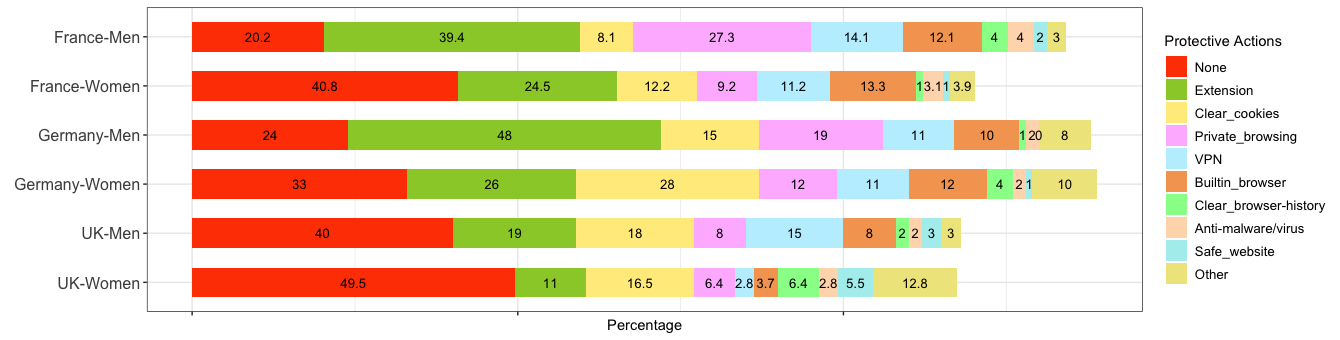} 
	\caption{Reported \emph{actions} across gender \& countries (x-axis shows \% in each gender-country group with n=100 approx.)\\
	\emph{Note:} The first (red) portion of the bars refer to `none', that is the \% of participants not taking protective actions.
	}
 \label{fig:actions}
 \end{figure*}
 
\subsection{Protective Actions Named by Participants}
\label{sec:protective_actions_self_report}
We investigate RQ2, that is ``What tracking protective actions do individuals employ online?", given their gender and country differences.
From the free-form text entered by participants (in responding to the questionnaire detailed in Section~\ref{sec:questionnaire_experience-action}), we identified all mentions of PETs or other protection strategies.

We summarise the \% participants naming each protective action, across gender and country in Figure~\ref{fig:actions}.
The noticeable patterns are that: (1) more women report to take no protective action (coded as `none') in all countries; 
(2) more UK participants report to take no action compared to those from Germany and France, across both gender; and
(3) the use of browser extension (coded as `extension') is the most named action for men across all countries, whereas UK and German women prefer clearing/rejecting cookies (coded as `clear\_cookies').

$34.7\%$ responses were categorised as `\textbf{none}' as participants wrote `none' or responded with text such as UK7 \emph{``I'm not aware of how I can protect myself (I haven’t knowingly allowed it)"}, or from GE48 \emph{``nothing although i dont like Companys [sic] having so much data about me its really not that bothering to me so i just keep on using the internet as before"}.

$65.3\%$ of participants named at least one protection method.
We list the type of protective method and provide example responses.
For \textbf{extension to browser}: responses included FR93 \emph{``I have installed Ublock Origin and added personalized precautions, and also Privacy Badger. Firefox also helps you with the cookies, fingerprint blocking,..."}.
For \textbf{reject, limit or delete cookies}: UK144 responded \emph{``None in particular other than declining the installation of cookies when they are mentioned."} or UK114 \emph{``I always opt out of marketing requests and clear cookies"} or FR138 \emph{``not much. I delete the cookies often, but that's it."}
For \textbf{private browsing/browser}: FR61 responded \emph{``Private browsing or onion routine (tor)"} and UK92 \emph{``i use brave browser and ..."}.
For \textbf{VPN (Virtual Private Network)}:  responses included FR92 \emph{``I have a vpn, and when a website asks for my identity..."}.
For \textbf{builtin browser options}: responses included GE78 \emph{``My browser (firefox) blocks most third-party tracking scripts"}. We note that this protective method category refers to mainstream browser options rather than browsers built for privacy as in the `private  browsing/browser category'.
For \textbf{anti -malware, -virus}: responses included UK3 \emph{``antimalware in my antivirus software and also keeping everything updated [sic]"}.
For \textbf{Clear browser history}: FR15 responded \emph{``Every day I delete my browser history, all of it..."}.
For \textbf{Safe website:} UK87 responded \emph{``not visited web pages that have come up as suspicious"}.

Participants also mentioned strategies that we grouped under `\textbf{other}'. These included actions such as: \emph{``I do not add my locations to social networking sites"} (UK79), 
\emph{``I try to avoid public wifi unless absolutely necessary"} (FR148), 
\emph{``avoid giving personal information, avoid social networks..."} (FR96) and
\emph{``I mostly use my work laptop and rely on it to have a good security software"} (GE174).

\begin{RedundantContent}
$38.4\%$ participants reported to use only $1$ type of protection method, $18.4\%$ reported $2$ different types, $4.9\%$ reported $3$, and $3.6\%$ reported at least $4$.
Approximately $16.6\%$ of participants reported `reject, limit or delete cookies' as tracking protection method.
Of these, $45\%$ relied only on rejecting, limiting or deleting cookies for tracking protection, whereas $55\%$ used other methods in addition to `reject, limit or delete cookies' (corresponding to approximately $9\%$ of the overall sample).
Approximately $49\%$ of participants reported tracking protection methods that did not involve `reject, limit or delete cookies'.

\textbf{Demographic Comparison}
We investigate RQ3a and RQ3b.
We summarise the protection methods named, and their use across countries and gender in Figure~\ref{fig:user_actions}. 
We observe significant differences between countries 
with more UK participants taking no protective action (`none', $p=.001$) or using a safe website ($p=.020$),
less UK participants using a browser extension ($p<.001$) or private browsing ($p=.003$), and
more DE participants clearing cookies ($p=.012$).
Between men and women overall, we observe significant differences with more men using a browser extension ($p<.001$), private browsing/browser (p=$.006$), VPN (p=$.007$), and women naming other actions (p=$.002$) or not taking any actions (p=$.006$). 
\end{RedundantContent}

\subsection{Association of Feelings with Protective Actions}
\label{sec:association_feelings_actions}
We investigate RQ3, that is ``How are individuals' feelings about third-party tracking associated with their protective actions?"
We answer this RQ in two steps: first we provide an overall descriptive view of all the feeling tones together with the proportion of participants naming the different actions for each feeling tone, in Figure~\ref{fig:all_feelings-actions}.
Second we investigate the statistical association between feelings tones and protective actions via a multivariate analysis, producing a spatial map in Figure~\ref{fig:map_feelings-actions}. 
We note that in contrast to the spatial map (in Figure~\ref{fig:map_feelings-actions}) that plots the strength of association between feelings' action profiles and actions' feelings profiles, Figure~\ref{fig:all_feelings-actions} provides a descriptive view of the raw data. 

\textbf{Overall View:}
Figure~\ref{fig:all_feelings-actions} shows that for all expressed feeling tones, some participants responded to take no protective action, under `none'. However, `none' accounts for a high proportion of action types for feelings in  the `okay/indifferent' categories (while also noting the smaller number of participants naming these feelings). 

\begin{RedundantContent}
  \begin{figure*}[h]
	\centering
	\includegraphics[keepaspectratio, width=\textwidth]{./figures/Rplot_feelings-actions2} 
	\caption{Overall view of feelings (x-axis) and actions (y-axis). Each point corresponds to the count of participants naming both the particular feeling and the particular action. The x-axis shows feeling tones from \emph{not\_okay} to \emph{distrust} that correspond to the \emph{generally not okay/negative} category, tones from \emph{ambivalence} to \emph{okay\_protected} corresponding to the \emph{sometimes okay, sometimes not okay category}, tones from \emph{okay} to \emph{none} referring to the \emph{generally okay/indifferent category} and \emph{should\_regulate} and \emph{not\_aware} for the \emph{other} category.}
 \label{fig:all_feelings-actions}
 \end{figure*}
 \end{RedundantContent}
 
\begin{figure*}[h]
	\centering
	\includegraphics[keepaspectratio, width=.98\textwidth]{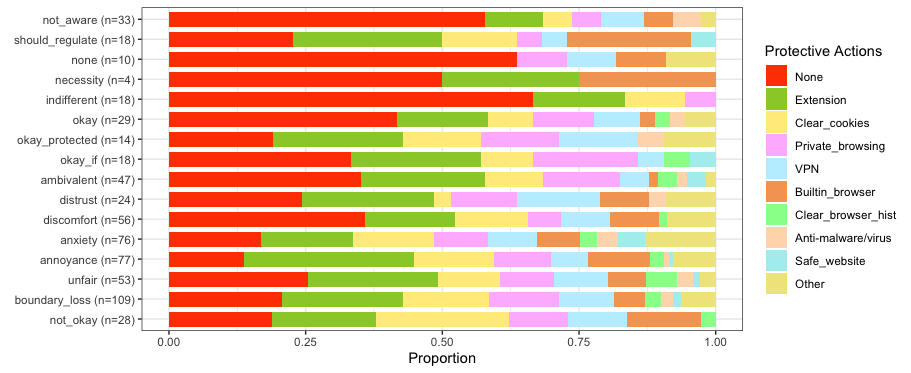} 
	\caption{Overall view of feelings and actions, with x-axis showing the proportion of participants naming the different actions for each feeling tone. 
The y-axis shows feeling tones from \emph{not\_okay} to \emph{distrust} that correspond to the \emph{generally not okay/negative} category, tones from \emph{ambivalence} to \emph{okay\_protected} corresponding to the \emph{sometimes okay, sometimes not okay} category, tones from \emph{okay} to \emph{none} referring to the \emph{generally okay/indifferent} category and \emph{should\_regulate} and \emph{not\_aware} for the \emph{other} category.}
 \label{fig:all_feelings-actions}
 \end{figure*}
 
\textbf{Multivariate Analysis:}
 \begin{figure*}[h]
    \centering
    \includegraphics[keepaspectratio,width=.85\textwidth]{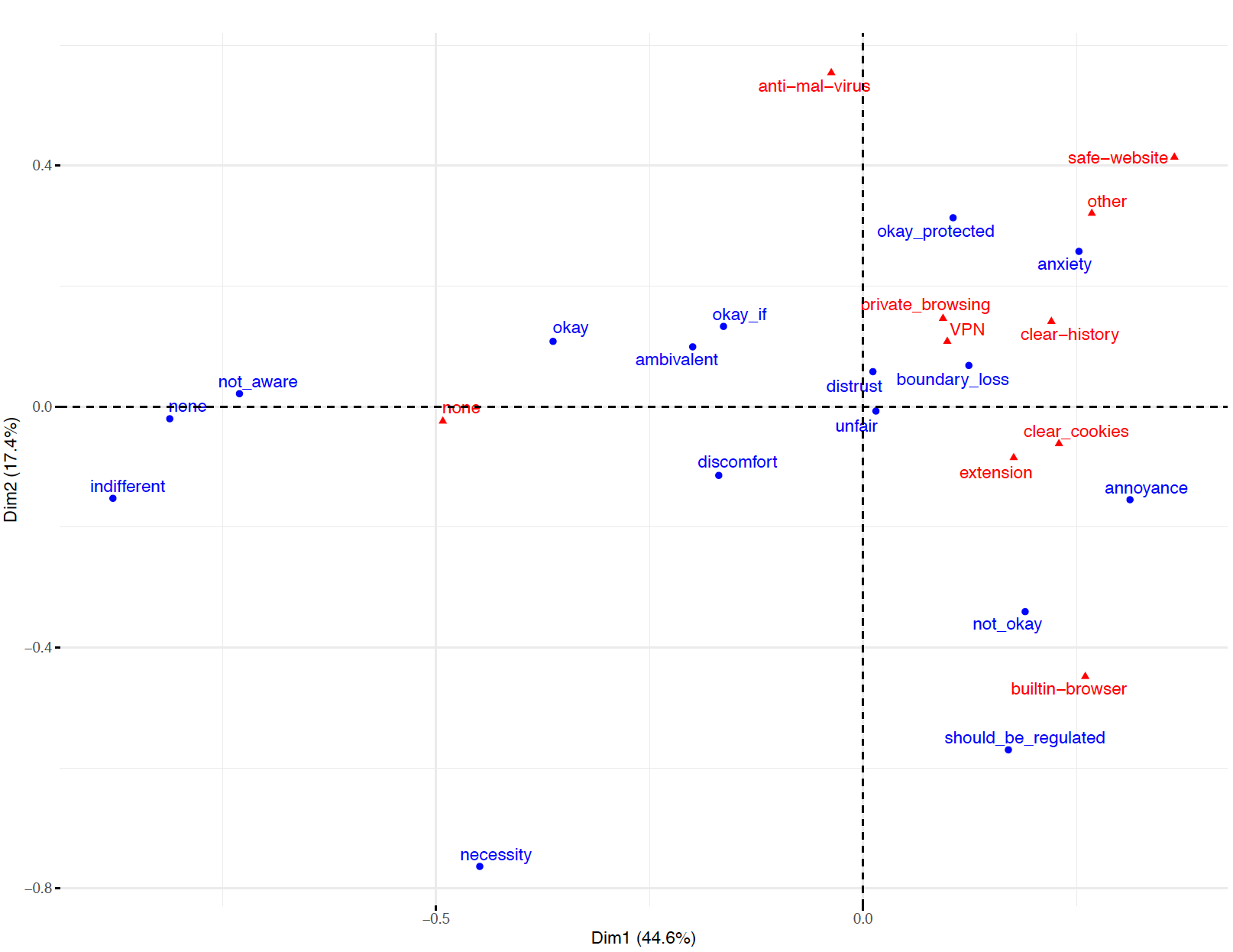}
    \caption{Spatial Map of Association between Feelings \emph{(in blue)} and Protective Actions \emph{(in red)}}
    \label{fig:map_feelings-actions}
\end{figure*}
We conduct a multivariate analysis of our dataset via a Correspondence Analysis (CA)~\cite{greenacre2017correspondence}, to investigate and visualise the relationship between expressed feelings and their protective action profiles. 
Our dataset is a contingency table of $16$ rows (feelings expressed with regards to TPT) and $10$ columns (protective actions named). 
In the following paragraphs, we report on the CA and results via the following steps (1) we compute the CA, (2) we visualise the spatial plot, and (3) we interpret the dimensions.

First, we compute the CA.
We find a significant association between the row and column variables with $\chi^2 = 165.634$, $p=.037$, and
the first dimension accounts for $44.58\%$ of the variance in the data while the second dimension accounts for $17.36\%$. Together these two dimensions account for $61.94\%$ of variability.

Second, we visualise the results via a spatial map, as provided in Figure~\ref{fig:map_feelings-actions}.
The spatial map shows the row and column profiles simultaneously in common space, where the proximity of the points demonstrates their similarity.
For example, expressed feelings (blue points) that are closer together show more similar protective action profiles compared to those that are further apart.

Third, we interpret the dimension 1 (the x-axis, Dim1) in Figure~\ref{fig:map_feelings-actions} as it intuitively shows a spectrum of feeling tones, and note that we do not notice a pattern for dimension 2.
Dim1 flows from the left with `no feeling' (indifferent, none, not aware),
to `being okay/mild feelings' (okay, ambivalent, okay if, discomfort), to
`strong emotional response' with regards to TPT (not okay, annoyance, anxiety).
The spatial map is intuitive in showing that the `no feeling' end of Dim1 (negative end of x-axis) is more closely associated with no protective action (none in red) and furthest to protective actions (such as using clearing cookies, using an extension or builtin browser options).
This corresponds to Figure~\ref{fig:all_feelings-actions} where the `indifferent', `none' and `not aware' tones have highest proportions of no protective actions.
In comparison, the `mild feeling' section of Dim1 is closer to, and the `strong emotional response' end (positive end of x-axis) is most closely associated with, employing these protective actions.
To summarise, (1) the feelings of being indifferent to TPT, having no feeling, being not aware of TPT, or viewing TPT as a necessity are furthest from action points;
(2) feeling okay under certain conditions (okay, okay if, ambivalent) are nearer to action points; whereas
(3) annoyance, anxiety emotional responses, as well as not okay, okay because one uses protection, call for regulation and feeling of boundary loss are nearest to protective action points.

\subsection{How Feelings Predict Protective Actions}
We investigate RQ4, that is ``How do individuals' feelings about third-party tracking predict whether they take protective actions or not, given their gender and country differences?"
We compute a mixed-effect binomial logistic regression with random intercept, with dependent variable `taking protective action' versus `not taking protective action' (as elicited from participant self-reports), and predictors gender, country and emotions.
The model is constructed as follows: $action \sim (1 | Participant) + Gender + Country + Feeling Tones$.
Feeling tones refer to the list of coded tones, except `not\_aware' which does not indicate any feeling.
We test the model with all tones versus with only negative tones, and find that the model with only negative tones has a better quality than the one with all tones, as shown by its lower Akaike's Information Criteria (AIC) and Bayesian Information Criteria (BIC). We therefore present the model $action \sim (1 | Participant) + Gender + Country + Negative Feeling Tones$.
This model performs significantly better than an intercept-only baseline model with $\chi^2(10) = 60.303, p<.001$.
It has a good fit ($C=0.701$), model accuracy of $70.6\%$ and $R^2$ of $13\%$. 
Table~\ref{tab:regression} reports that men were twice more likely to take protective actions than women, with odds ratio ($OR$) $=2.05, p<.001$, that German and French participants were approximately twice more likely to take protective actions than UK participants, with $OR=1.83, p=.006$ and $OR=1.78, p=.008$ respectively. In addition, participants who reported feeling tones of `not\_okay', `boundary\_loss' (invaded), annoyance or anxiety, also showed higher likelihood of taking protective actions (between $OR=2.53$ and $OR=3.84$), compared to those not naming these feelings.

\begin{table}
\centering
\caption{Binomial logistic regression for taking protective actions versus not.}
\label{tab:regression}
\footnotesize
\resizebox{\columnwidth}{!}{
\begin{tabular}{l|rrrr@{}l}
\toprule
& \textbf{Est.}  & \textbf{OR}  & \textbf{95\% CI} & \multicolumn{2}{c} {\textbf{p-value}}\\
\midrule
(Intercept) & -0.71 & 0.49 & [0.32 - 0.75 & .001&*** \\ \midrule
men (vs women) &0.72 &2.05& [1.43 - 2.95]&<.001&***\\ 
\midrule
Germany (vs UK) & 0.60&1.83&	[1.19 - 2.81] &.006&** \\
France (vs UK)& 0.58 &1.78 &	[1.16 - 2.74] &.008&** \\
\midrule
not\_okay: true (vs false) &1.19 & 3.28 & [1.30 - 8.27] &.012 &*\\
\midrule
boundary\_loss: true (vs false) & 0.93& 2.53 &[1.49 - 4.30] &.001& ***\\
\midrule
unfair: true (vs false)& 0.60 &1.83 & [0.95 - 3.50] &.070 \\
\midrule
annoyance: true (vs false)  &  1.35& 3.84& [2.00 - 7.37] &<.001&***\\
 \midrule
anxiety: true (vs false) &1.31&3.72&[1.98 - 6.99]	&<.001&***\\
  \midrule
discomfort: true (vs false) & 0.41 &1.50&[0.80 - 2.81] &.206 \\
\midrule 
distrust: true (vs false) &0.73 & 2.08 & [0.82 - 5.26]& .121\\ 
\bottomrule
\end{tabular}
}
\footnotesize{\emph{Note: Significance codes of $‘***’.001, ‘**’ .01, ‘*’ .05 $}}\\
\end{table}

\section{Discussion}
This paper makes a mixed-methods contribution (qualitative and quantitative) to user-centred privacy research, in particular on tracking protection. 
The main take-aways of this paper are that: (1) the majority of individuals (71.8\%) have a negative feeling about tracking, (2) overall 34.7\% do not take a protective action, some of whom also feel negatively, (3) protective actions are closely associated with (and predicted by) particular negative feelings, gender and country, 
and (4) there are indications of a gender gap and country differences in feelings and protective actions.
In this section, we first discuss our findings in relation to literature and related work. Second, we discuss the implications for different stakeholders and offer recommendations and paths for future work.

\subsection{Our findings in the wider research context}
\emph{Feelings:} While previous research did not find a clear causal link between individuals' cognitive evaluation of tracking (that is their understanding and mental models) and their protective behaviour~\cite{mathur2018characterizing},
our findings support positions that feelings can provide useful inputs to judgments, decisions and behaviour when individuals' cognitive evaluation about a situation or event is in-accurate or incomplete~\cite{damasio2008descartes}. 
This study therefore improves on cognitive-oriented investigations of tracking and user privacy. 

Our participants' negative feelings include strong language and are spread across nuances of feeling tones, and in particular not only cover a perception of privacy threats, such as being monitored, of unfair practices and emotional responses of anger, anxiety, distrust or discomfort, but also the experience of a deep invasion to their personal realm.
These findings augment previous findings of users' perception of privacy-invasiveness of tracking~\cite{ur2012smart, mcdonald2010beliefs,turow2009americans}.
We also complement previous research~\cite{ur2012smart,mcdonald2010beliefs,auxier2019americans}, with some of our participants expressing other feelings, such as sometimes being okay with tracking, being okay under certain conditions, or being overall okay.

\emph{Protective Actions:} 
In the context of the type of actions following particular negative feelings~\cite{lerner2001fear}, 
we notice some slight differences in protective behaviour, albeit without clear distinctions, between annoyance and anxiety (Figure~\ref{fig:all_feelings-actions}). 
However, we note the significant prediction of actively taking protective actions, by
annoyance, anxiety or feeling invaded, compared to feeling discomfort, distrust or unfairness.
In addition, while previous research reported individuals desisting from using services, such as location-based services, or taking retributive actions, such as issuing complaints, following anger and anxiety~\cite{jung2018investigation}, our study is limited in not capturing these behaviours. 
However, of the $34.7\%$ of participants reporting to take no protective action, the ones feeling negatively about tracking may be adopting passive coping methods such as acceptance or avoidance, as previously noted in relation to privacy threats~\cite{park2020users,jung2018investigation,cho2020privacy}.



\emph{Gender influence:} Our findings provide indications of a `privacy gender gap', complementing previous research on gender differences on concerns~\cite{baruh2017online,sheehan1999investigation,youn2009determinants}, where 
women in our study expressed more negative feeling tones combined, compared to men, who expressed more ambivalence or `okay under certain condition' tones compared to women -- yet men are twice more likely to take any protective actions, aligning with previous findings on protective behaviour~\cite{sheehan1999investigation,redmiles2018net}.
This suggests support for three decades old feminist critiques of the unequal nature of privacy online -- that privacy is not equally on the side of women as it is for men~\cite{allen1988uneasy,mackinnon1989toward}.
More research is needed into how these differences play out online and how to empower women 
to exercise their desired type and level of privacy.





\emph{National influence:}
We find that the British are less active in protective actions against tracking, and provided less reports of negative feelings, 
where action differences 
aligns to previous work~\cite{coopamootoo2020usage}.
Protective actions may relate to the effects of tracking being less clear or being less equipped to act compared to other modalities in the advertising ecosystem, where  for example, a higher \% of UK participants reported to object to direct marketing (i.e. being contacted directly via email or text messages) in the 2019 Eurobarometer survey, compared to those from Germany or France~\cite{eurobarometer2019european}.
The British may also be less expressive or possibly have higher trust in authority mandated privacy protection.
This phenomenon, with the national culture as an underlying cause, needs further investigation. 
Additionally, the impact of the depth of experience of privacy violation (as expressed by German and French's stronger language for boundary loss) on protective actions would benefit investigation.

\subsection{Implications \& Recommendations}
We discuss the implications for different stakeholders, provide recommendations and highlight avenues for future work. 

\emph{Users:} Individuals' negative emotions, feelings of unfairness and deep invasion depict the dis-empowered reality of users, where protective action are also not necessarily effective.
We recommend users to consider privacy-oriented browsers (such as the Brave browser, as chosen by some of our participants) over other options. 
In addition, to facilitate the development of more privacy-empowered online communities, specially supporting certain user groups, users may also be encouraged to openly share about their experiences of privacy issues and protection methods online, 
so as to facilitate the support of social and trusting connections on the feelings/experience--protective actions link.

\emph{Educators, privacy technology designers \& providers:} to understand who are more receptive (given their feelings) to privacy technologies, and to customise ways to up-skill individuals given their characteristics of gender and country.
Other ways to improve privacy practice include free online courses, from reliable sources such as the national data protection authorities.
We recommend that educators and privacy-enhancing technology (PET) designers make it clear what protection is offered by particular PETs, in a language that address users' concerns as expressed via their feelings about tracking, irrespective of their mental models. 
It would also be helpful to establish PETs repositories, and make vetted recommendations more accessible to the lay user. Existing lists for the general public include that of the Electronic Frontier Foundation’s~\cite{EFF_PETs}
or the European Agency for Cyber Security’s (ENISA)~\cite{ENISA_PETs}.

\emph{Researchers:} to deepen knowledge into the factors between particular negative feeling and protective action, such as awareness of and the obstacles to using protective methods and the support needed, across gender and country characteristic, or whether particular feeling-tones activate or inhibit actions.
This includes research into 
different coping behaviours.
In addition, 
individuals likely cope with tracking threats based on the combined judgement of cognition and affect, while influenced by factors such as perceived self-efficacy and response-efficacy -- future work to include these variables, as well as 
awareness of the limitations of protection methods, where favoured tracking protection methods such as browser extensions, may not effectively block ads and trackers~\cite{pujol2015annoyed}. 
We also recommend research into diverse gender identities and user groups, in particular to understand their privacy experiences, vulnerabilities and challenges, and in comparison to the already available literature focused on cisgender. 
Furthermore, tracking issues also exist in platforms such as apps and IoT devices, and 
across platforms. 
While some efforts have been made for example in Apple's recent App Tracking Transparency (ATT) policy \cite{AppleTT}, it comes with its own issues such as increasing the number of privacy prompts and
its effectiveness in improving user privacy remains as research questions. 


\emph{Regulators and national data protection authorities:} 
to set guidelines for actual fair practices, such as to avoid dark patterns that nudge users’ acceptance despite their feelings and concerns, and for companies and service providers to demonstrate these fair attributes to customers, 
and to provide fair and inclusive practices for diverse user groups.
It would also be helpful for researchers and designers to work with regulators and authorities, in preparing user-centric guidelines that support the deployment of new privacy technologies. 

\section{Conclusion}
This paper adds to literature with an understanding of individuals' experiences of tracking and protection practices online.
It adds to the mostly cognitive-focused literature of user privacy by providing novel findings on how feelings associates with and predicts protective actions.
It describes a mixed methods approach, including (1) elicitation and synthesis of individuals' feelings and their own description of their protective actions, and
(2) quantitative analyses and visualisation of associations.
It discusses the findings in the context of previous work and what the findings mean for various privacy stakeholders
and make recommendations.
As highlight for future work, it proposes 
that although particular feelings about tracking are closely linked to protective actions, further research is needed into connecting these feelings with effective action, while gender/country differences point to needing customised methods and support for accessible protection.


\section*{Acknowledgements}
This research was supported by a Newcastle University research fellowship.
We would like to thank our shepherd, Simone Fischer-Hubner, and the Usenix Security 2022 reviewers for their feedback which helped to improve the paper.

\bibliographystyle{plain}
\bibliography{repository,repository1,repository2,privacy,emotion}
\begin{appendix}
\onecolumn
\section{Codebook}
\label{sec:ap_codebook}
\begin{table*}[h]
    \centering
    \footnotesize
    \caption{Codebook of Feelings with regards to Third-Party Tracking (with example words mentioned in responses)}
    \label{tab:feelings_codebook}
    \resizebox{.7\textwidth}{!}{
        \begin{tabular}{ll}
            \toprule
            \textbf{Emotion}                 & \textbf{Explanation \& Example words in responses}                                        \\
            \midrule
            Not Okay                         & Participants say `not okay', `not like'                                                   \\
                                             &                                                                                           \\
            Boundary loss 	& There is a sense of loss of boundaries wrt to info privacy, or beyond info boundaries              \\ 
                                             & E.g words: invasion of privacy, tracked,unsafe, naked, raped, exposed, invaded, \\
                                             & watched, violated, exploited, insecure \\
                                             &                                                                                           \\
            Unfair practice                  & A sense of unfair actions from others                                                     \\
                                             & E.g. words: used, misused, sold, manipulated, ill-informed, not you making decisions,     \\
                                             & no chance to decide, info stolen, betrayed, dishonest, sneaky, sleazy, immoral,           \\
                                             & illegal, like a rat-test subject. No control, trapped, cheated, obscure, powerless, unethical        \\
                                             &                                                                                           \\
            Annoyance                        & Expressed anger-related emotion                                                           \\
                                             & E.g. words: Annoyed, angry, disgust, irritated, exasperated, bothered                     \\
                                             &                                                                                           \\
            Anxiety                          & Expressed worry-related emotion                                                           \\
                                             & E.g. words: Scared, worry, cautious, creepy, spooky, unnerving, concern                          \\
                                             &                                                                                           \\
            Distrust                         & Expressed suspiciousness                                                                  \\
                                             & E.g. words: Suspicious, not trust, wary, uncertain                                                   \\
                                             &                                                                                           \\
            Discomfort                       & Expressed discomfort or overwhelm                                                         \\
                                             & E.g. words: uncomfortable, overwhelm, unsettled, unpleasant, uneasy, disturbed            \\
                                             &                                                                                           \\
            Should be regulated              & Expressed that third-party tracking should be regulated/protected                                          \\
                                             & E.g. words: should be regulated, should be protected, should be banned,                    \\
                                             & should be made aware, should be transparent \\
                                             &                                                                                           \\
            Okay                             & Expressed that they were overall okay with tracking or used specific words.               \\
                                             & E.g. words: I am okay, I am alright                                                       \\
                                             &                                                                                           \\
    Ambivalent                       & Expressed that they are sometimes okay sometimes not okay, fine for this reason           \\
                                             & but not fine of other reason                                                              \\
                                             &                                                                                           \\                          
            Okay, protected                  & Expressed that they were okay with tracking because they protect themselves anyway.       \\
                                             & E.g. words: I am okay as/because I protect myself.                                        \\
                                             &                                                                                           \\
            Okay, if                         & Expressed that they would be okay with tracking IF ….                                     \\
                                             & E.g. words: I am okay if they do that / …if I know / …but have to inform                  \\
                                             &                                                                                           \\
            Necessity                        & E.g. words: it’s a necessity/a must, impossible for normal user to do                     \\
            &                                                                                           \\
            Not aware                        & E.g. words: I don’t know enough, I don’t know                                             \\
            &                                                                                           \\
            Indifferent                      & E.g. words: indifferent, insensitive, I don’t care                                        \\
            &                                                                                           \\
            None                             & Participants say that they feel nothing                                                   \\
            \bottomrule
        \end{tabular}
    }
\end{table*} 

\end{appendix}

\end{document}